\documentclass{aastex}

\usepackage{graphicx}
\usepackage{amssymb}
\usepackage{natbib}
\usepackage{amsmath}
\usepackage{url}

\newcommand{\myemail}{mat@igam.uni-graz.at}

\shorttitle{Analytic model of a Moreton wave}
\shortauthors{Temmer et al.}

\begin{document}

\title{Analytic modeling of the Moreton wave kinematics}

\author{M. Temmer}
\affil{IGAM/Kanzelh\"{o}he Observatory, Institute of Physics, Universit\"at Graz, Universit\"atsplatz 5, A-8010 Graz, Austria}\email{\myemail}

\author{B.~Vr\v{s}nak and T.~\v{Z}ic}
\affil{Hvar Observatory, Faculty of Geodesy, University of Zagreb, Ka\v{c}i\'{c}eva 26, HR-10000, Croatia}

\author{Veronig A.M.}
\affil{IGAM/Kanzelh\"{o}he Observatory, Institute of Physics, Universit\"at Graz, Universit\"atsplatz 5, A-8010 Graz, Austria}

\begin{abstract}
The issue whether Moreton waves are flare-ignited or CME-driven, or a
combination of both, is still a matter of debate. We develop an analytical model describing the evolution of a large-amplitude coronal wave emitted by the expansion of a circular source surface in order to mimic the evolution of a Moreton wave. The model results are confronted with observations of a strong Moreton wave observed in association with the X3.8/3B flare/CME event from January 17, 2005. Using different input parameters for the expansion of the source region, either derived from the real CME observations (assuming that the upward moving CME drives the wave), or synthetically generated scenarios (expanding flare region, lateral expansion of the CME flanks), we calculate the
kinematics of the associated Moreton wave signature. Those model input parameters are determined which fit the observed Moreton wave kinematics best. Using the measured kinematics of the upward moving CME as the model input, we are not able to reproduce the observed Moreton wave kinematics. The observations of the Moreton wave can be reproduced only by
applying a strong and impulsive acceleration for the source region expansion acting in a piston mechanism scenario. Based on these results we propose that the expansion of the flaring region or the lateral expansion of the CME flanks is more likely the driver of the Moreton wave than the upward moving CME front.
\end{abstract}


\maketitle

\section{Introduction}

Solar flares and coronal mass ejections (CMEs) are explosive processes that are able to generate
large-scale wave-like disturbances in the solar atmosphere \citep[e.g.][]{warmuth07}. Signatures of
such disturbances were first imaged in the hydrogen H$\alpha$ spectral line and called Moreton
waves after \citet[][see also Moreton \& Ramsey, 1960]{moreton_orig60}. Typically, Moreton waves
appear as propagating dark and bright fronts in H$\alpha$ filtergrams and dopplergrams, respectively, which can be attributed to a compression and relaxation of the chromospheric plasma. The disturbance propagates with a speed in the order of 1000~km~s$^{-1}$ \citep[e.g.][]{moreton60,zhang01,warmuth04a,veronig06}, which led to the conclusion that such a phenomenon cannot be of chromospheric origin, but is the surface track of a coronal
disturbance compressing the underlying chromosphere \citep[sweeping-skirt hypothesis;
see][]{uchida68}. Moreton waves are generally observed to be closely associated with the flare
impulsive phase \citep{warmuth04a}, which often coincides also with the acceleration phase of
the associated CME \citep[cf.][]{zhang01,vrsnak04b,maricic07,temmer08}.

Moreton waves are observed to propagate perpendicular to the magnetic field, and the initial magnetosonic Mach numbers are estimated to lie in the range of \mbox{M$_{\rm ms}\sim$1.4--4}, suggesting that they are at least initially shocked fast-mode waves \citep{narukage02,narukage04,warmuth04b}. In their late propagation phase the wave perturbations undergo a broadening, weakening, and deceleration until \mbox{M$_{\rm ms}\sim$1} is reached. These results indicate that Moreton waves are a consequence of shocks formed from large amplitude-waves that decay to ordinary fast magnetosonic waves, which is in line with the flare initiated ``blast wave'' scenario \citep[e.g.,][]{warmuth01,khan02,narukage02,vrsnak02a,hudson03,narukage04}. Further evidence for the close association to shocks is the quasi-simultaneous appearance of Moreton waves and radio type II bursts, which are one of the best indicators of coronal shocks \citep[e.g.,][]{khan02,pohjolainen01,pohjolainen08,warmuth04b,vrsnak05b,vrsnak08}.

Wave-like disturbances were for the first time imaged directly in the corona by the EIT instrument aboard the Solar and Heliospheric Observatory (SoHO), thereafter called EIT-waves \citep[][]{moses97,thompson98}.
They were considered to be the coronal manifestation of the Moreton wave \citep{thompson99}, but statistical studies revealed discrepancies in their velocities. EIT waves were found to be two to three times slower than Moreton waves \citep{klassen00}. Today, their relation to Moreton waves and the generation mechanism of EIT waves is very much debated \citep[e.g.,][]{delannee99,wills-davey99,chen00,biesecker02,cliver04,warmuth04b,cliver05,vrsnak05b,chen06,attrill07,veronig08}.

In the present paper, we solely focus on Moreton waves, which are generally accepted to be a chromospheric response to coronal shock waves. In particular, we study their generation mechanism and address the issue whether they are flare-ignited or CME-driven, or a combination of both, which is still a matter of debate. To this aim, we developed a simple analytical model which describes the launch and propagation of Moreton waves. (Note that the presented model does not intend to evaluate generation mechanisms which may cause EIT waves.) We use for the model different input parameters acting as source that drives wave, first derived from CME observations (assuming that the upward moving CME drives the wave), and second using synthetically generated scenarios (to emulate alternative driving mechanisms). By confronting the results derived from the model with observations we aim to find constraints on the possible drivers of the wave. For this we use the outstanding observations of the Moreton wave associated with the X3.8/3B flare-CME event from January 17, 2005. We emphasize that the event was characterized by a very distinct and fast Moreton signature, indicating that it was caused by a coronal fast-mode shock \citep[c.f.][]{warmuth04b}.

The observations of the Moreton wave and associated CME and flare under study are presented in Sect.~2. The model is described in Sect.~\ref{model}. The results are given in Sect.~4. Discussions on the results, constrains of model input parameters by observations, and final conclusions are presented in Sect.~5.

\section{Observations}

Associated with the January 17, 2005 3B/X3.8 flare event, a fast Moreton wave starting at $\sim$09:44~UT was observed with high time cadence ($\lesssim$1~min) in full-disk H$\alpha$ filtergrams at Kanzelh\"ohe Observatory. The wave propagated at a mean velocity of 930~km~s$^{-1}$ up to a distance of 500~Mm from its source location \citep[for more details on the wave measurements and its propagation characteristics we refer to][]{veronig06}. The flare and its associated coronal mass ejection (CME) occurred at [N15,W25]. From this region actually two fast CMEs were launched within short time and in our study we are focusing on the second event. The Large Angle and Spectrometric Coronagraph \citep[LASCO;][]{brueckner95} instrument C2 aboard the Solar and Heliospheric Observatory (SoHO) imaged the first CME at 09:30~UT and the second CME at 09:54~UT. The linear plane-of-sky speed of the first CME was of $\sim$2100~km~s$^{-1}$ and of the second CME of $\sim$2500~km~s$^{-1}$ as observed with LASCO C2 and C3 \citep[LASCO catalogue;][]{yashiro04}. The study is performed over the time range 09:30--09:54~UT, hence, in the interval of interest we assume no impact on the CME kinematics due to the possible merging process with the previous event. The early CME evolution could be observed with the GOES12 Soft X-ray Imager \citep[SXI;][]{hill05}. Rising CME loops could be identified in 9~SXI frames with high time cadence \citep[$\sim$2--4~min; see][]{temmer08}.

After co-aligning the GOES/SXI and H$\alpha$ observations, the distances of the CME leading edge as well as the Moreton wave fronts were measured using as null-point the wave ``radiant point"\footnote{Note that in \cite{temmer08} the Sun-center was used as null-point for the distance measurements of the CME.}, which was derived from circular fits to the earliest observed wavefronts \citep[for details see][]{veronig06}. From running ratio SXI images the height-time profile of the erupting CME structure was measured.

\begin{figure}
\epsscale{0.7} \plotone{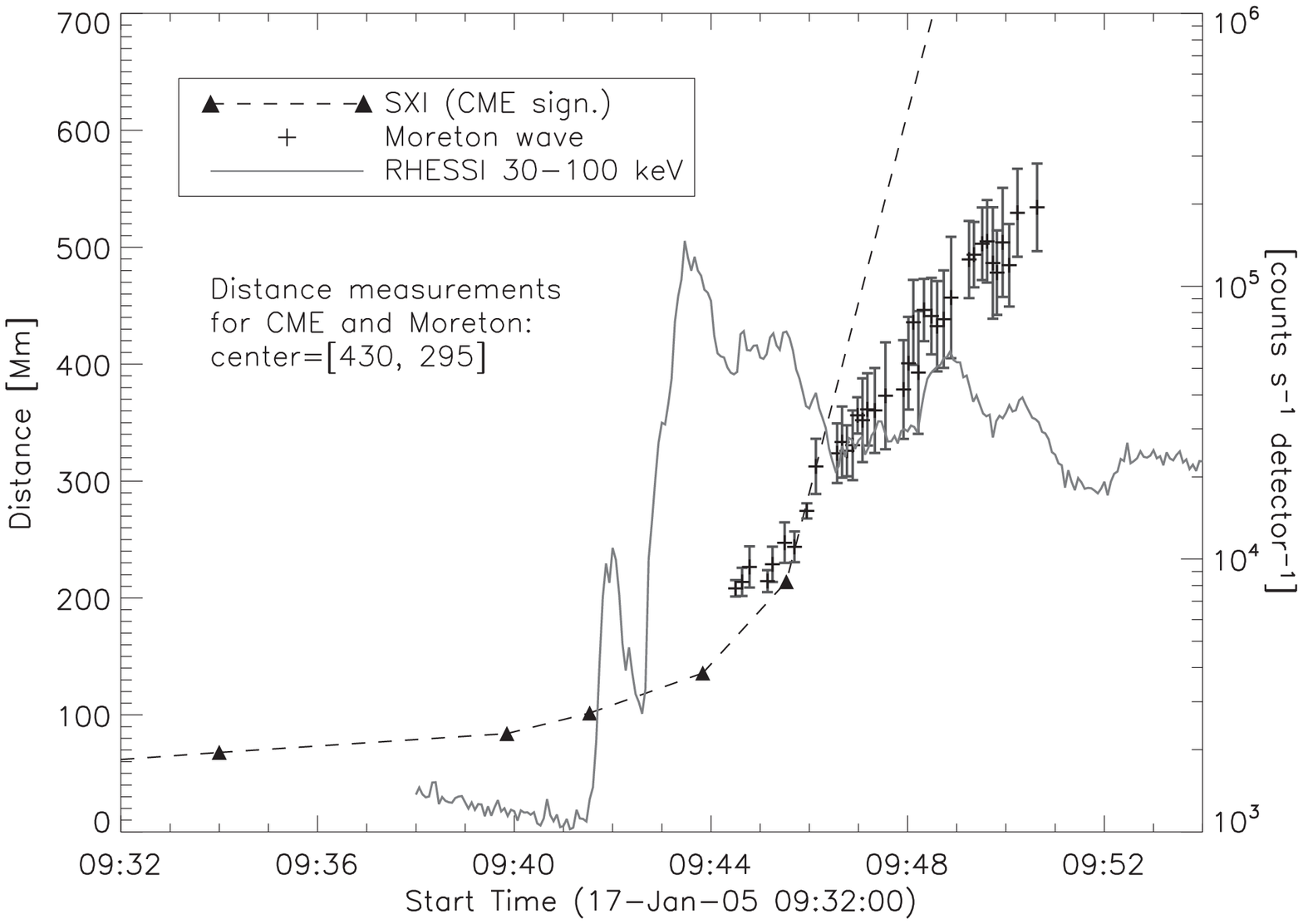}
 \caption{Evolution of the flaring process, CME take-off, and Moreton wave. Triangles indicate the measured CME height as derived from SXI images. Gray lines give the RHESSI flare hard X-ray flux for the energy band 30--100 keV. Plus signs give the kinematics of the observed Moreton wave front together with error bars. Around $\sim$9:42~UT we observe the first H$\alpha$ flare brightenings, the first HXR peak, as well as the CME and Moreton wave initiation (derived from back extrapolation). }
 \label{cme-kin}
\end{figure}

In Fig.~\ref{cme-kin} we show the propagation of the Moreton wave together with the associated CME during its initial phase up to $\sim$1~R$_{\odot}$ and the flare hard X-ray (HXR) flux measured with the Ramaty High-Energy Solar Spectroscopic Imager \citep[RHESSI;][]{lin02} in the non-thermal energy range 30--100~keV. From the second derivative of the height-time measurements we determined the onset of the CME fast acceleration phase, i.e. the launch time of the CME, at $\sim$09:40--09:42~UT. The back-extrapolated Moreton wave as well as the first HXR burst started at $\sim$9:42~UT \citep[see][]{veronig06}. The CME acceleration reached its peak of $4.4\pm0.3$~km~s$^{-2}$ at $\sim$09:46~UT, and ends at $\sim$10:06~UT \citep[cf.][]{temmer08}. For the full CME kinematics up to 30~R$_{\odot}$ we refer to \cite{vrsnak07} and \cite{temmer08}.

A composite dynamic radio spectrum for that day over the frequency range 600~MHz--20~kHz combining Artemis, DAM and WAVES measurements can be found under \url{http://secchirh.obspm.fr/select.php}. The radio signatures show a rather complex situation most probably due to the launch of two CMEs for which a detailed study is given by \cite{bouratzis09}. Associated with the event under study was a metric type II radio burst at 09:43--09:46~UT reported from San Vito, Italy (SVTO; spectral range 70--25~MHz) and also from Learmonth, Australia at 09:44--09:47~UT (LEAR; spectral range 65--25~MHz) as reported from the Solar Geophysical Data (SGD) under {\it Solar Radio Spectral Observations} (\url{ftp://ftp.ngdc.noaa.gov/STP/SGD/}). Both stations report shock velocities of 1500~km~s$^{-1}$ using a one-fold Newkirk model which is consistent with an MHD shock moving through the solar corona. A group of type III bursts occurred 09:41--09:47~UT, matching the main RHESSI peak.

\begin{figure}
 \epsscale{0.7} \plotone{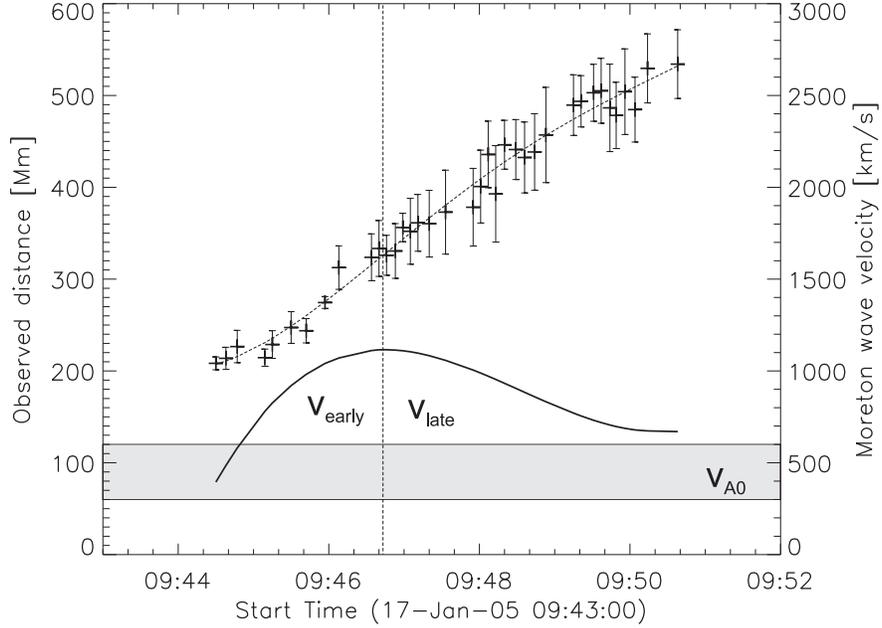}
 \caption{ Evolution of the Moreton wave kinematics (left scale) and velocity evolution (right scale). Plus signs with error bars show the Moreton wave front together with a 4$^{th}$ order polynomial fit (dashed line). The solid line shows the derivative of the polynomial fit, i.e.\ the velocity profile. The velocity curve can be divided into an early phase of increasing velocity ($v_{\rm early}$) followed by a later phase of decreasing velocity ($v_{\rm late}$) with the inflection point at $\sim$09:47~UT (dashed vertical line). The gray horizontal bar indicates the local Alfv\'en velocity $v_{\rm A0}$ for the corona outside of active regions that lies in the range 300--600~km/s.}
 \label{wave-kin}
\end{figure}

In Fig.~\ref{wave-kin} the distance-time and velocity-time profile of the observed Moreton wave is shown. The profile shows an increase in velocity with an initial speed of 400~km~s$^{-1}$ until it reaches a maximum speed of 1100~km~s$^{-1}$ at $\sim$09:47~UT, afterwards the velocity decreases. This temporal behavior can be interpreted as nonlinear evolution of the wavefront. First, the wavefront steepens until a discontinuity appears, i.e. the shock formation starts. Then follows a phase of shock amplitude growth, which is reflected in shock acceleration and intensification \citep[see Figures 4 and 5 in][]{vrsnak00a}. Finally, after the shock amplitude attains its maximum, the wave gradually decays to an ordinary fast-mode wave \citep[cf.][]{zic08}.

\begin{figure}
\epsscale{0.7} \plotone{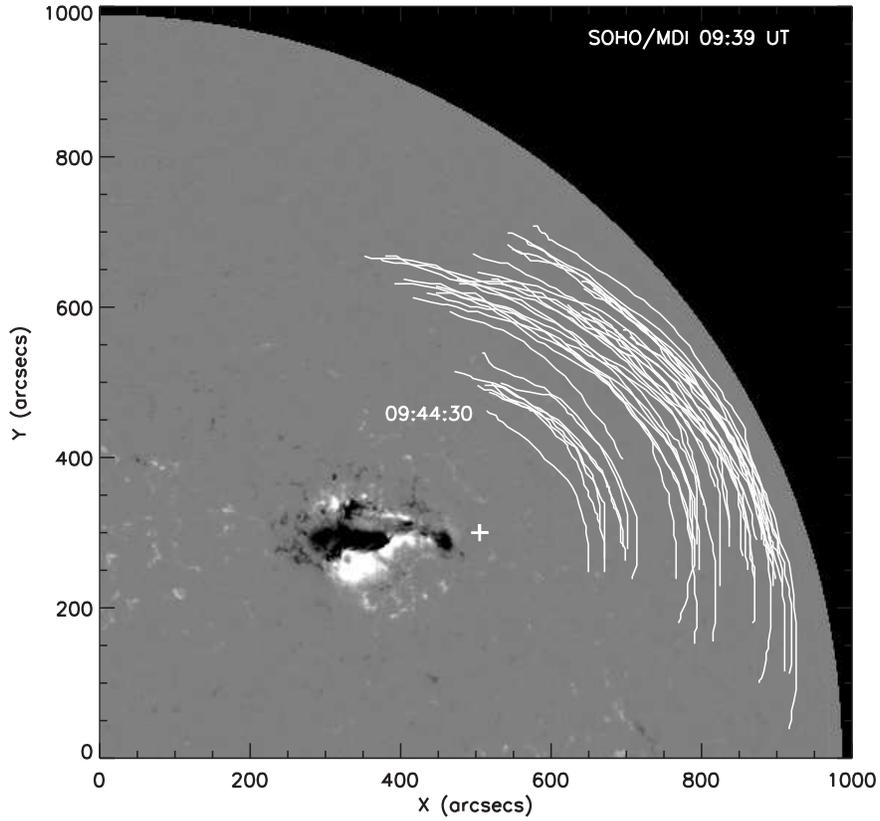}
\caption{ SOHO/MDI magnetogram scaled to a magnetic field strength of $\pm$700~G. Solid white lines show the Moreton wave fronts observed in H$\alpha$ images. The first wave front at 09:44:30~UT is clearly located outside the area of strong magnetic fields. Plus sign indicates the wave radiant point derived from a circular fit to the earliest observed wave front \citep[for details see][]{veronig06}.}
 \label{mdi}
\end{figure}

Fig.~\ref{mdi} shows the derived Moreton wave fronts with respect to the photospheric magnetic field. The first wave appearance is clearly located outside the active region. Since the wave propagated well outside the active region, Alfv\'en speeds for the corona can be considered to lie in the range of 300--600~km~s$^{-1}$ \cite[e.g.][]{narukage02,warmuth05}. The high velocity of the wave within a low Alfv\'en speed environment as well as  the associated metric type II radio burst suggest that the wave is at least initially shocked \citep[e.g.][]{gopalswamy98,klassen99}.

The main criteria derived from the observations which our model results have to meet are 1) general kinematics of the wave, 2) velocity evolution and 3) timing of the shock formation.

\section{The model}\label{model}

We would like to emphasize that the following analytical model is kept as simple as possible and can thus only reproduce the general characteristics of the propagation of the disturbance. The model will simulate the Moreton wave by applying a driver which is a circular source region that may expand and move translatory at the same time.

Three types of source expansion are applied following the terminology by \cite{vrsnak05a}: 1) The radius of the source is kept constant, i.e.\ there is no expansion of the source in time during its upward motion. Accordingly, plasma can flow behind the driver and the source acts as blunt-body driving a bow-shock. 2) The source radius expands with a constant radius-to-height ratio, $r(t)/h(t)$, acting as a combined bow-shock/piston driver. 3) The source expands only in lateral direction without upward motion and plasma can not flow behind the contact surface, according to which the driver acts as piston mechanism.

Our first intention is to investigate whether the Moreton wave could be produced by the upward moving CME, using the height-time measurements derived from the CME observations as input for the expanding source. We consider this model input for scenarios where the source acts as bow-shock and combined bow/piston driver for the wave (different strengths and proportions between the upward motion and lateral expansion of the driver are applied). Our second intention is to emulate an expanding flare region or the lateral expansion of the CME flanks for which we use synthetic expansion profiles. Such kind of model input is considered for a source that acts as piston driver mechanism for the wave. The results from the model will be compared to the kinematics of the January 17, 2005 Moreton wave to estimate what kind of source expansion reproduces the general characteristics of the observed wave kinematics best.

\begin{figure}
 \epsscale{0.7} \plotone{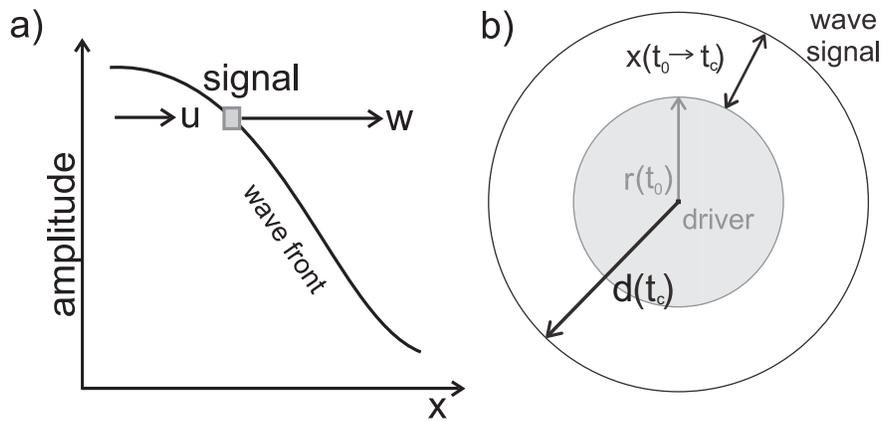}
 \caption{a) Definition of the term ``signal'': a given element (gray box) of the disturbance profile is referred to as signal. It is characterized by the propagation speed $w$ and the associated plasma flow velocity $u$. b)
The signal radius $d$ at time $t_{\rm c}$, $d(t_{\rm c})$, is defined as the sum of the source size $r$ at the time $t_{\rm 0}$ when the signal was launched, and the propagated distance $x$ of the signal up to the time $t_{\rm c}$ (see Equ.~\ref{radius}). }
 \label{sig}
\end{figure}

We suppose that the source accelerates to a high velocity, which causes a large amplitude coronal
disturbance that is capable of compressing the underlying chromosphere to produce the Moreton
wave. The term large-amplitude waves should emphasize that the wave evolution can not be described
through linearized equations. For more details on the terminology of large scale waves we refer to
\cite{vrsnak05a} and \cite{warmuth07}. In the case of a large amplitude wave, the rest frame velocity $w$ of a given wavefront element (hereinafter called ``signal'') depends on two quantities. First, it
depends on the local magnetosonic speed $v_{\rm ms}$, which is larger than in the unperturbed
plasma due to the plasma compression, and is thus related to the perturbation amplitude. Second, it
must be taken into account that a given signal propagates through a moving plasma, since the plasma
flow velocity $u$ associated with the perturbation amplitude is not negligible (see
Fig.~\ref{sig}a). Consequently, the rest frame velocity of the signal equals to $w=v_{\rm ms}+u$
\citep[see][]{landau87}, i.e., elements of larger amplitude propagate faster. Due to the nonlinear
evolution of the wave front, its profile steepens and after a certain time/distance a discontinuity
forms, marking the onset of shock formation \citep[][]{landau87,mann95_s,vrsnak00a,zic08}.

Generally, the dependence of $v_{\rm ms}$ on the perturbation
amplitude cannot be expressed straightforwardly. However, in the
case of a low plasma-to-magnetic pressure ratio $\beta$ which is assumed here, the
relationship simplifies, since the Alfv\'en velocity $v_{\rm A}$
is much larger than the sound speed, and under the frozen-in condition
in the case of perpendicular wave propagation, the plasma density $\rho$ is
proportional to the magnetic field strength $B$, i.e.\ $v_{\rm ms}\approx
v_{\rm A}\propto \sqrt{\rho}$. \cite{vrsnak00a} have shown that in
such a situation the relationship between the local propagation
speed and the amplitude becomes very simple: the local value of
$v_{\rm A}$ can be expressed as $v_{\rm A}=v_{\rm A0}+u/2$, where
$v_{\rm A0}$ is the local Alfv\'en velocity in the unperturbed plasma.
Bearing in mind that $w=v_{\rm A}+u$, one finally finds that the
wave element propagates at the rest frame speed $w=v_{\rm A0}+3u/2$ \citep{vrsnak00a}.

Since the phase velocity of the signal depends on its amplitude $u$ and the ambient Alfv\'en
velocity $v_{\rm A0}$, the evolution of the wavefront depends on the spatial
distribution of $v_{\rm A0}$ and the evolution of the amplitude. The simplest possible situation is
propagation of the wave in a medium where $v_{\rm A0}$ is uniform. In such a case, the phase
velocity changes only due to the amplitude evolution, which is governed by the energy conservation.
For example, in the case of a spherically symmetric source, creating a spherically symmetric wavefront (Fig.~\ref{sig}b), the amplitude is inversely proportional to the distance $d$, i.e., decreases as $d^{-1}$, whereas in the cylindrical symmetry it decreases as $d^{-1/2}$ \citep{landau87}. Note that in the case of freely-propagating shock waves (blasts), the amplitude decreases also because the leading edge of the perturbation (having the
highest velocity) propagates faster than the low-amplitude segments in the trailing edge. This causes perturbation profile broadening, which must be compensated by an amplitude decrease
\citep{landau87}.\footnote{Note that in a medium where the Alfv\'en velocity decreases steeply enough with the
distance, the leading edge might be slower than the trailing edge. In such a case,
the wavefront slows down, whereas the amplitude increases.}

Of course, in the solar corona the Alfv\'en velocity is far from being uniform. Even if the coronal
structural inhomogeneities are neglected, it changes with height and depends on the distance from
active regions \cite[e.g.,][]{warmuth05}. In such a situation, where the spatial distribution of $v_{\rm
A0}$ is generally unknown, one has to investigate the wavefront kinematics by calculating
the amplitude evolution for various reasonable spatial distributions of $v_{\rm A0}$.
However, instead of this, we apply an analogous procedure, where we take $v_{\rm A0}$
uniform, and describe the signal amplitude and the phase-velocity evolution by different functional
forms. In other words, instead of presuming a function that describes the change of $v_{\rm A0}$
with distance $d$ from the wave source, we directly presume a function that describes the wave
evolution. In particular, we use the power-law function
\begin{equation}\label{pl}
  f(d)= d^{-\alpha}\,
\end{equation}
and exponential function
\begin{equation}\label{expon}
  f(d)= {\rm e}^{-d/p}\,.
\end{equation}
Applying different decay lengths (denoted in the power-law function by $\alpha$ and in the exponential by $p$) we can reproduce a weak or strong attenuation of the signal. Note that $f=1$ would
represent a plane wave without decay as achieved for $p\rightarrow\infty$ and $\alpha\rightarrow0$.
On the other hand, large $\alpha$ or small $p$ represents a strong attenuation.

Beside the power-law and exponential function, we also employ as a kind of reference, the
functions:
\begin{equation}\label{cylin_dec}
  f(d)= \frac{1}{\sqrt{d}}\,
\end{equation}
and
\begin{equation}\label{spher_dec}
  f(d)= \frac{1}{d}\,,
\end{equation}
which describe the amplitude decrease of cylindrically and spherically symmetric sound waves,
respectively.

The initial amplitude of a given signal is determined by the velocity of the source surface
$v_{\rm s}$. At the starting time $t_{0}$ when the signal is launched, $u(t_0)=v_{\rm s}(t_0)$,
since the flow velocity has to be equal to the contact-surface velocity. The geometry of the source
is considered as a radially expanding surface of cylindrical (arcade expansion) or spherical shape
(volume expansion) with a radius $r(t)$ centered at the height $h(t)$. Applying the Huygens-Fresnel
principle, one finds that due to the presumed symmetry of the source and the presumed homogeneity
of the ambient plasma, the wavefront elements are concentric with the source surface (cf.
Figs.~\ref{sig}b,~\ref{f1} and~\ref{f1b}).

We follow the signals which are emitted continuously from the source surface for the time span
$t_{0}$ until a certain time $t_{\rm c}$ at each small time step $\Delta t=t_{i}-t_{i-1}$. The distance $x$ traveled by the signal from the time $t_0$ when it was emitted, until the time $t_i$, is calculated iteratively. Using the expression

\begin{equation}\label{radius}
x(t_{i}) = x(t_{i-1}) + \left( v_{\rm A0} +  v_{\rm
s}(t_{i-1})~\frac{3}{2}~f(t_{i-1})  \right) \Delta t \, ,
\end{equation}
we obtain the distance from the source region center, $d(t_{i})=r(t_0)+x(t_{i})$, where $r(t_{0})$ is the radius of the source surface at the time $t_0$, when the signal was emitted (Fig.~\ref{sig}b). Note that $x(t_0)=0$ and $d(t_{0})=r(t_{0})$, and Eq.~\ref{radius} has to be integrated from $t_{0}$ to $t_{\rm c}$.

Considering the mimicked Moreton wave as the extension of the outermost signal measured at the solar surface
(cf.\ arrows in Figs.~\ref{f1} and~\ref{f1b}), we derive for each time step $\Delta t$ the
propagation of the wave as distance $d_{\rm M}(t)$. Hereinafter, this outermost signal that is
considered to mimic the Moreton wave, will be denoted as the \textit{ground track signal} (GTS).

\section{Implementation and interpretation of the model}\label{implement}

In the following, distance-time plots and velocity profiles are shown for the propagated GTS resulting from our model. The results are confronted with the observed Moreton wave kinematics. Due to the huge spectrum of possibilities obtained by varying and combining the different model parameters, we will show here only representative model results, i.e.\ those which match the observational criteria of the Moreton wave best. The successful model will reproduce the general characteristics of the observed Moreton wave in terms of 1) kinematics, 2) velocity evolution (increasing velocity until $\sim$09:47~UT followed by decreasing velocity), and 3) shock formation around the onset of the type II burst ($\sim$09:43~UT), i.e., before or close in time to the first appearance of the Moreton wave ($\sim$09:44~UT).

The wave-like disturbance that generates the Moreton wave is assumed to propagate approximately near the coronal base. Under this assumption, the value for $v_{A0}$ lies in the range of $\sim$300--600~km~s$^{-1}$ \cite[][]{warmuth05}. To ease the comparison between the model results and the observations (bearing in mind also other aspects of the CME/flare event) we use for the model the absolute time in UT. The parameter $t_{0}$ varies around $\sim$09:42~UT which is close to the onset of the fast acceleration stage of the CME and the flare onset in H$\alpha$ and HXRs. The parameter $t_{\rm c}$ is the time at which the Moreton wave was observed the last time \cite[$\sim$9:54~UT; see][]{veronig06}. The time range $t_{0}$--$t_{\rm c}$ is subdivided into time steps $\Delta t$=10~s, i.e.\ each 10 seconds the position of the wavefront and the GTS is calculated.

\subsection{Model results based on observed CME kinematics}

\begin{figure}
\epsscale{0.7} \plotone{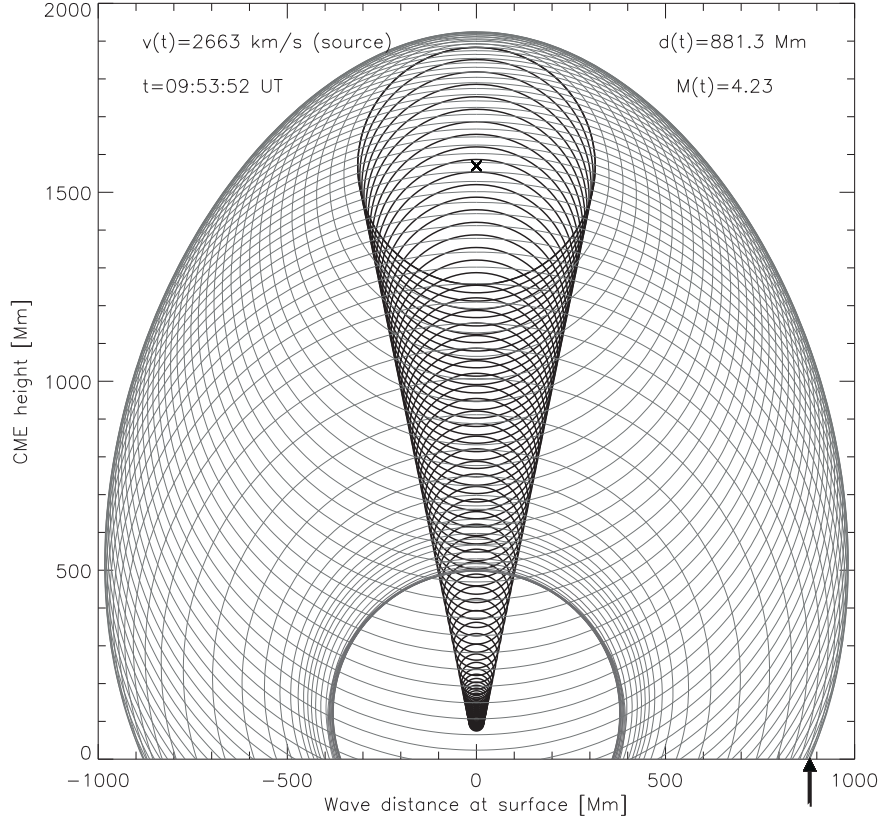}
 \caption{ Black circles show the upward moving CME (observed CME kinematics) which is expanding during its upward movement from $t_{0}$ until $t_{\rm c}$ with $r(t)/h(t)$=0.2 (combined bow/piston scenario). The cross indicates the height of the CME center at $t_{\rm c}$. Gray circles show the evolution of calculated signal radii launched from the circular source surface (CME) followed in steps of $\Delta t$=10~s. For time $t_{\rm c}$=9:53:52~UT the outermost signal measured at the surface $h=0$, i.e. the mimicked chromospheric Moreton wave front, is marked with an arrow. }
 \label{f1}
\end{figure}

In Fig.~\ref{f1} we give a snapshot of the propagated signals (circles) that were emitted during
the upward motion (along the $y$-axis) of an expanding source. The kinematics for the upward
moving source is taken from the CME observations, and the type of source expansion acts as a
combined bow shock/piston driver for the emitted signals with $r(t)/h(t)$=0.2, i.e.\ source size is
proportional to height at each time $t$. The decay of the signal is based on a cylindrical
geometry of the source (see Equ.~\ref{cylin_dec}). The first signals are emitted at
$t_{0}$=9:41:52~UT when the CME had a height of $h(t_0)$=105~Mm and an initial size of
$r(t_0)$=21~Mm. The surrounding Alfv\'en speed of the unperturbed plasma is chosen as $v_{\rm
A0}$=400~km~s$^{-1}$. From $t_{0}$ on, we follow the signals every 10~s, until they have reached a
certain extension at $t_{\rm c}$=9:53:52~UT (Fig.~\ref{f1}). Note that signals
which are launched right after $t_0$ have the longest time to evolve, signals launched close to
$t_{\rm c}$ the shortest. At 9:53:52~UT the CME has a height of $h(t_{\rm c})$=1570~Mm and a size
of $r(t_{\rm c})$=314~Mm. The arrow in Fig.~\ref{f1} indicates the propagated distance $d_{\rm M}(t_{\rm
c})$=881~Mm of the GTS, i.e.\ the mimicked Moreton wave at 9:53:52~UT.

\begin{figure}
 \epsscale{0.6} \plotone{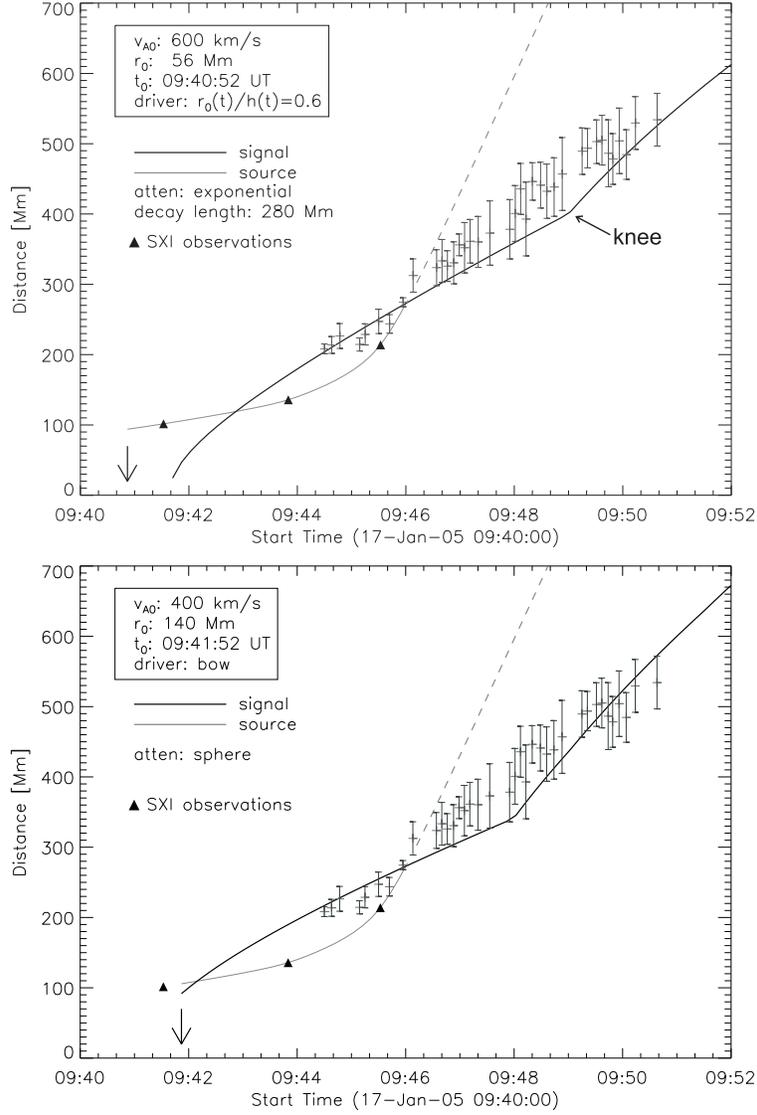}
  \caption{Calculated distance-time profile of the generated Moreton wave signal (solid black line)
using observed CME kinematics as input for the model. The different parameter values used are specified
in the legend. The solid gray line is the kinematics of the driver of the wave signal, i.e.\ in the present case
the CME, the dashed gray line represents the stage when the driver is no more directly related to the evolution of the wavefront profile. The triangles indicate the CME front
measurements from GOES/SXI observations. The kinematics of the observed Moreton wave
front is plotted by plus signs with error bars. Top panel: a combined bow/piston driven
scenario is applied using a ratio $r(t)/h(t)$=0.6 for the increase of the source size during
its upward movement. Bottom panel: a bow shock driven scenario is applied, i.e.\ keeping the source
size constant ($r(t)$=140~Mm) during its upward movement. The arrow pointing to the $x$-axis
indicates the launch time of the first signal $t_{0}$.}
 \label{cme-real}
\end{figure}

Fig.~\ref{cme-real} shows the calculated GTS distance versus time using the observed CME kinematics as
input for the upward moving source for two different types of the source expansion. The top panel
of Fig.~\ref{cme-real} is supposed to mimic a combination of a bow shock and piston driven scenario;
the source was expanding during its upward motion self-similarly with a constant
ratio of $r(t)/h(t)=0.6$. The bottom panel of Fig.~\ref{cme-real} supposes the source to act as a
rigid-body driver, i.e. the radius was kept constant during its upward movement with $r(t)$=140~Mm, imitating a bow-shock scenario.

The derived kinematics of the GTS show a distinct feature of a ``knee'' as indicated in the top panel of Fig.~\ref{cme-real}. The feature occurs when a later emitted GTS passes the preceding one\footnote{In the specific case of our model the overtaking GTS was launched when the source speed changed from subsonic to supersonic.}, i.e.\ the
knee marks the time of the shock formation \citep{vrsnak00a}.

\begin{figure}
\epsscale{0.7} \plotone{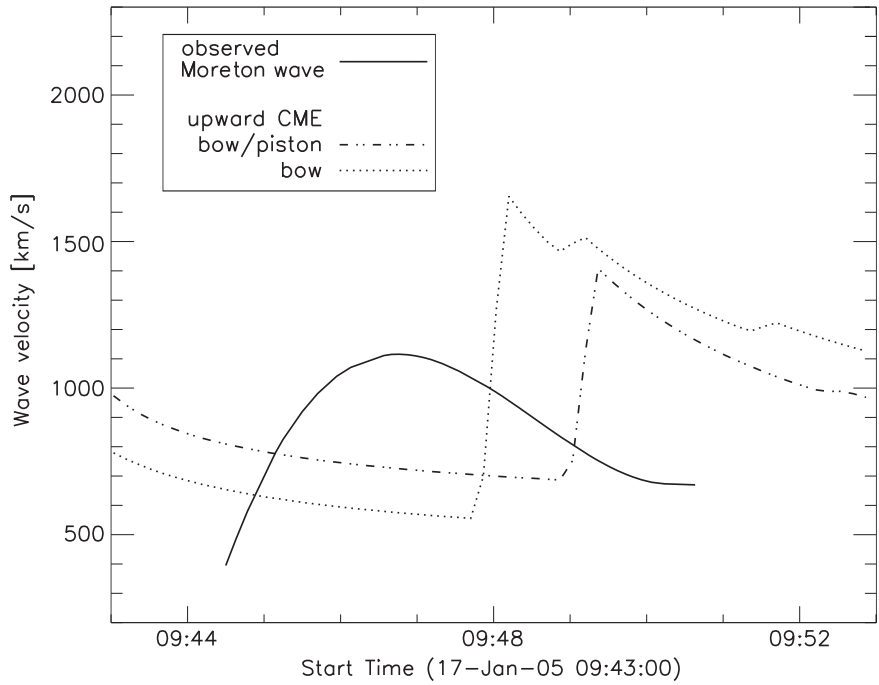}
 \caption{ Velocity profiles derived from the generated wave kinematics shown in Fig.~\ref{cme-real} using the observed kinematics of the upward moving CME as input. A combined bow/piston scenario (dashed-dotted line) and a bow shock scenario (dotted line) is applied for the driver. For comparison, the velocity profile of the observed Moreton wave is plotted as solid line.}
 \label{cme-real-vel}
\end{figure}

\begin{figure}
\epsscale{0.7} \plotone{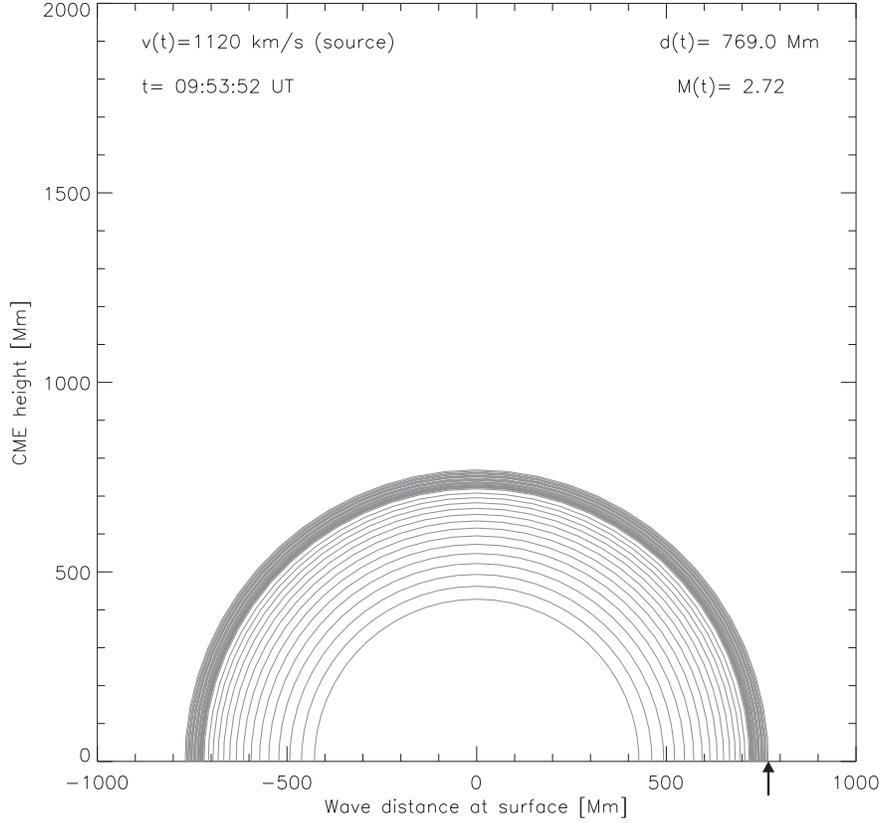}
 \caption{Disturbance signals emitted from an expanding piston source using $t_a$=400~s and $a$=2.8~km~s$^{-2}$. The arrow indicates the intersection of the outermost signal with the $x$-axis, i.e.\ the propagated way of the mimicked Moreton wave at 9:53:52~UT. The initial source size is $r(t_{0})$=140~Mm.}
 \label{f1b}
\end{figure}

From Fig.~\ref{cme-real} it can be seen that the first phase of the observed Moreton wave could be
partly mimicked but not its later evolution. The knee, which represents the time of the shock formation, occurs $\sim$4--6 minutes after the first Moreton wave front was observed. In Fig.~\ref{cme-real-vel} the according velocity profiles are plotted for the scenarios presented in Fig.~\ref{cme-real}. For both scenarios, CME acting as combined bow/piston and bow driver, the GTS is of decreasing velocity until $\sim$09:47~UT and the velocity of the GTS at $\sim$09:51~UT (last observational data point) is about 1.5 times as high as for the observed Moreton wave. Hence, the CME is a too fast driver which generates a too fast GTS at large distances. Although various kinds of parameter values were applied, it was not possible to reproduce the general observational characteristics of the Moreton wave. From this we conclude that, using a fast upward moving driver for the model, like the observed CME, all generated GTS profiles reveal 1) increasing velocity after $\sim$09:47~UT and 2) a shock formation several minutes after the first observed front of the Moreton wave (cf.~Fig.~\ref{wave-kin}), which is not consistent with the observations.

\subsection{Model results based on a synthetic kinematical profile of the source}\label{synth}

From the calculated GTS kinematics using real CME observations, it became clear that the radially
upward moving CME, imitating a bow or combined bow/piston scenario, cannot reproduce the observed Moreton wave characteristics. In order to investigate alternative driving mechanisms, we use as input parameters a synthetic kinematics of an expanding source acting as piston mechanism. As simplest approach, we assume that during the radial expansion the center of the source is fixed at the surface, i.e.\ $h(t)$=0, in order to imitate a spherical or cylindrical piston. The synthetic kinematics consists of an acceleration phase $t_{\rm a}$ of constant acceleration $a$, until a certain velocity is reached by which the source expands further. This enables us to study the signal evolution emitted from very differently expanding driving sources, ranging from sudden impulsively to gradually accelerating.

\begin{figure}
\epsscale{0.7} \plotone{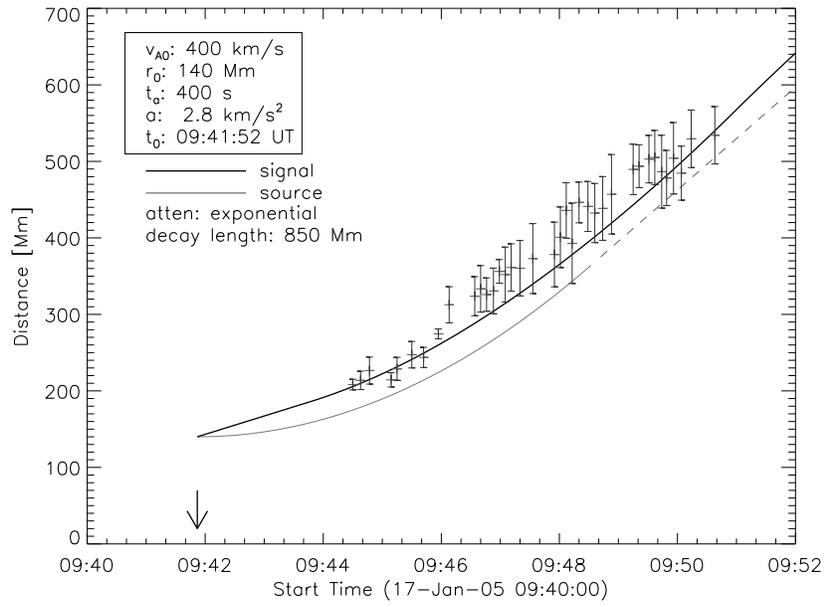}
  \caption{Same as Fig.~\ref{cme-real} but for a GTS derived from a synthetic kinematical profile of the source representing a spherical piston. The synthetic kinematical profile would be appropriate for a fast upward moving CME event. Different parameter values are specified in the legend.}
 \label{f5}
\end{figure}

In Figs.~\ref{f1b} and~\ref{f5} a relatively gradual expansion of a spherical piston is represented. We use as input an initial source size of $r(t_{0})$=140~Mm accelerating over a time span of $t_{a}$=400~s with $a$=2.8~km~s$^{-2}$ (final velocity 1120~km~s$^{-1}$). The arrow in Fig.~\ref{f1b} indicates the propagated distance $d_{\rm M}(t_{\rm c})$ of the GTS, i.e.\ the mimicked Moreton wave. The shock formation time was obtained at 09:48:32~UT, hence, several minutes after the first Moreton wave front was observed. The corresponding velocity profile (shown in Fig.~\ref{f7} as dashed line) reveals an increase of velocity of the GTS in the late propagation phase after 09:47~UT, although a strong decay (exponential) was applied to the GTS its kinematics. Similar to what we obtained applying the observed CME kinematics such an acceleration behavior of the source cannot mimic the observed Moreton wave. To compensate for the delayed timing of the shock formation, a shorter and more impulsive acceleration of the source expansion would be required to reproduce adequately the Moreton wave propagation.

\begin{figure}
 \centering
 \epsscale{0.6} \plotone{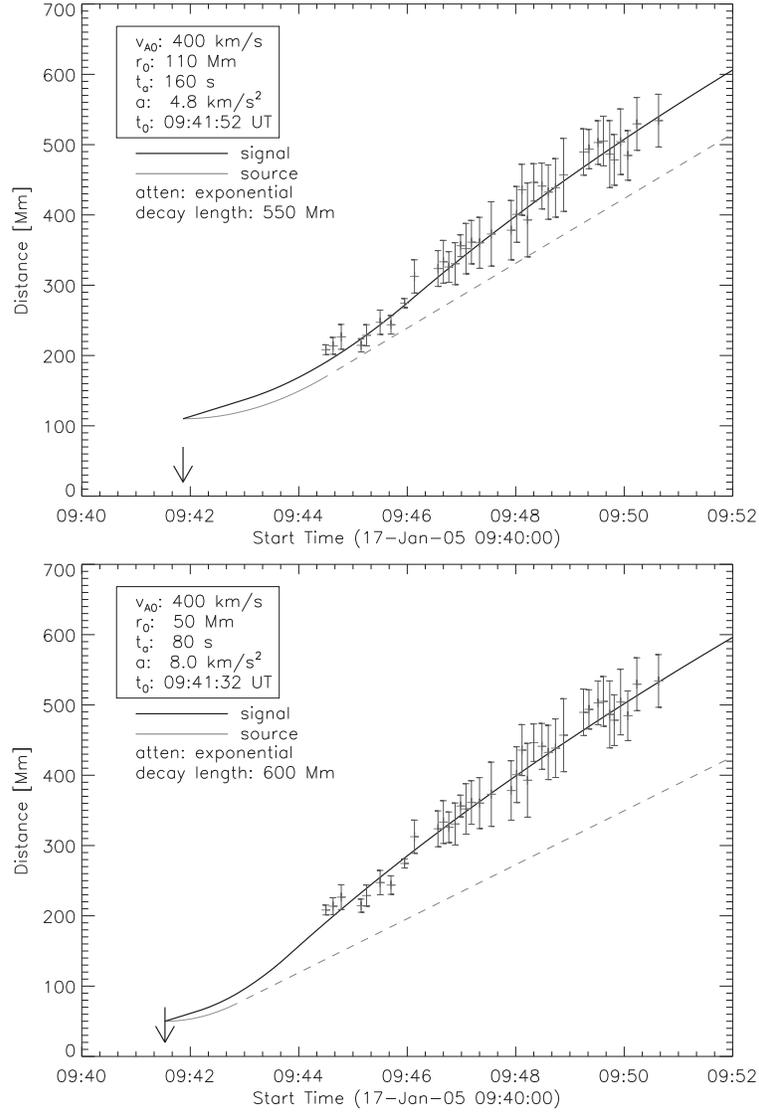}
  \caption{Same as Fig.~\ref{f5} but for synthetic kinematical profiles of different source sizes, acceleration times and strengths. Different parameter values are specified in the legends.}
 \label{f6}
\end{figure}

\begin{figure}
 \epsscale{0.7} \plotone{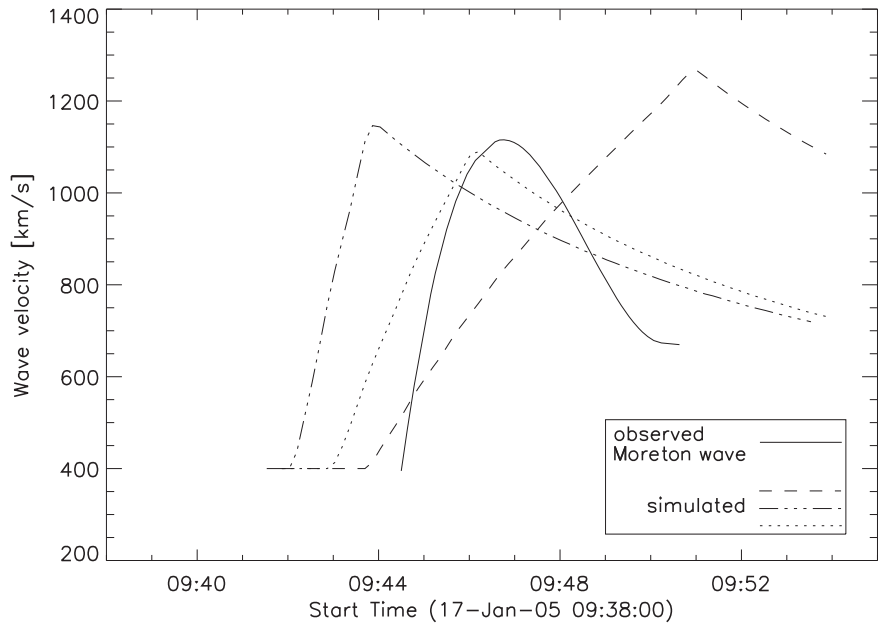}
   \caption{Velocity profiles derived from the simulated waves as shown in Figs.~\ref{f5} and~\ref{f6}. The dashed line represents the wave driven by a source region expansion of $a$=2.8~km~s$^{-2}$ and $t_{a}$=400~s. The dashed-dotted line represents the source region expansion of $a$=8.0~km~s$^{-2}$, $t_{a}$=80~s, whereas the dotted line shows the case $a$=4.8~km~s$^{-2}$ and $t_{a}$=160~s. For comparison the velocity profile of the observed Moreton wave is plotted as solid line.}
 \label{f7}
\end{figure}

The top panel of Fig.~\ref{f6} shows the expansion of a smaller source of $r(t_{0})$=110~Mm of a shorter and stronger acceleration ($t_{a}$=160~s; $a$=4.8~km~s$^{-2}$) in comparison to the previous scenario. The calculated GTS from this case shows a very good match with the observed Moreton wave kinematics as well as its velocity profile (dotted line in Fig.~\ref{f7}). The timing of the shock formation at 9:44:32~UT is close to the first detected Moreton wave front ($\sim$9:44:30~UT). Since after the shock formation the GTS propagates faster than the later emitted signals we assume that the source is acting only temporarily as piston. The time range during which the wavefront evolves independently from the driver is indicated as dashed gray line in Fig.~\ref{f6}. From this we derive, the source surface would need to expand from the initial size of 110~Mm up to 170~Mm to mimic the resulting Moreton wave (solid gray line in Fig.~\ref{f6}). The initial source size of $\sim$110~Mm would roughly correspond to the diameter of the active region (cf.~Fig.~\ref{mdi}). A further scenario is presented in the bottom panel of Fig.~\ref{f6} with source parameters comprising an initial size of $r(t_{0})$=50~Mm and a very impulsive expansion (short and strong acceleration) of $t_{a}$=80~s and $a$=8~km~s$^{-2}$. The synthetic kinematics of the calculated GTS matches the observed Moreton wave reasonably and the shock formation for this scenario takes place at 9:42:52~UT. The source surface, acting as a temporary piston, would need to expand from its initial size of 50~Mm up to 75~Mm. Considering the velocity profile (dashed-dotted line in Fig.~\ref{f7}) the GTS reaches its peak velocity before 09:47~UT, however, decreases very rapidly. Compared to the earlier scenario (source parameters: $r(t_{0})$=110~Mm; $t_{a}$=160~s; $a$=4.8~km~s$^{-2}$; marked with the dotted line in Fig.~\ref{f7}) the match is worse, however, still reasonable within the limits of such a simple model.

In Fig.~\ref{f7} we show the velocity profiles from the simulated wave kinematics as given in Figs.~\ref{f5} and~\ref{f6}, and compare them to the velocity profile derived from the observed Moreton wave (solid line). We obtain the best match for a wave which is assumed to be driven by a shortly and strongly accelerating source (dotted line); a more impulsive expansion of the source would generate a profile of comparable velocity at the last point of observation close to 09:51~UT, but peaks earlier (dashed-dotted line). Such source behavior could be interpreted as the expanding flanks of a CME or the volume expansion of a flare. On the other hand, a weak and long acceleration similar to the upward moving CME (dashed line) reveals substantial inconsistencies to the observed wave profile (late peak, final velocity too high).

\section{Discussion and Conclusion}

The analytical model presented here is based on tracing the evolution of a large amplitude wave. This is justified since Moreton waves are caused by a strong compression of the chromosphere (otherwise the wave would not be seen in H$\alpha$). There are several unknown factors whose implementation would be beyond the scope of this model. For example, we considered a homogeneous corona where the density and the Alfv\'en velocity do not change, neither in the vertical nor in the horizontal direction, taking $v_{\rm A0}$ in the range 300--600~km~s$^{-1}$. Recent observational studies showed that the magnetosonic speed $v_{\rm ms}$ (we assume $v_{\rm ms}\approx v_{\rm A0}$) can drop down to a local minimum of 300--500~km~s$^{-1}$ around the height $\sim$2~R$_{\odot}$ but then rises steadily up to a local maximum of $\sim$1000~km~s$^{-1}$ at a height between 3 and 4~R$_{\odot}$ \citep{mann03, warmuth05}. \cite{vrsnak04b} obtained from observations of type II bursts that on average the magnetosonic speed attains a local minimum of $v_{\rm ms}\approx$400~km~s$^{-1}$ around 3~R$_{\odot}$ and a broad local maximum of $v_{\rm ms}\approx$500~km~s$^{-1}$ in the range of 4--6~R$_{\odot}$. Besides, the previous CME event which started about 40~min earlier \citep[LASCO catalogue;][]{yashiro04} from the same active region might affect the actual value of the Alfv\'en velocity too.

Furthermore, we did not take into account the accurate relation between the plasma flow and source velocity $u$,
i.e.\ the CME velocity, but simply used a one-to-one relation. We approximated $u$ by the CME speed which is appropriate concerning the upper part of the moving and expanding CME but does not hold for the lateral direction, i.e., from which the GTS kinematics is determined. We tried to account for this by reducing the CME speed by $\sim$60\%, thus maintaining the CME kinematical profile as model input but with a lower speed. However, also that option did not result in a better match between the generated GTS and the observed Moreton wave.

An important factor for the derived model results is the decay factor used to attenuate the signal. Since in the corona the distribution of density $\rho(r)$, magnetic field $B(r)$, and Alfv\'en speed $v_{A0}(r)$ are unknown, we use different ``decay functions'' (see Equ.~\ref{pl}--\ref{spher_dec}). It had to satisfy two criteria: it should be strong enough to decelerate the signal in its late propagation phase but should not, due to its strength, delay the timing of the shock formation. We used geometry dependent factors adapted from sound waves (cylindrical and spherical), i.e., without implementing a magnetic field \citep[for details see][]{zic08}. Formal decay factors, like power-law and exponential functions, were used to put the decay to the limits either having no attenuation or very strong attenuation and to account for the unknown distribution of $v_{\rm ms}$. \cite{pagano07} investigated the role of magnetic fields for an expanding and upward traveling CME and showed that a spherical cloud without a magnetic field drives a wave that propagates to longer distances than that with a weak open field \citep[see Fig.~7 in][]{pagano07}. This implies that the presence of a magnetic field would result in a stronger signal decay than obtained from our simple approaches. Since we were not able to reproduce the wave using the limits for the decay factor (strong versus no attenuation), we suppose that even utilizing more sophisticated decay factors, the disturbance generated by the CME forehead would not be able to reproduce the observed Moreton wave.

Using the observed CME kinematics as input parameters the model could not reproduce the general characteristics of the observed Moreton wave. The timing of the shock formation (``knee'') was not appropriate but occurred later than the first observed Moreton wave front. The velocity profile was not conform and the final velocity was too high in comparison to the observed Moreton wave. By varying the initial source size as well as the behavior during the source evolution (bow, piston or combined bow/piston driver), the GTS kinematics was shifted to a larger or smaller propagation distance, however, the shock formation always appeared too late \citep[see also][]{zic08}. Similar results are obtained by applying different start times for the signal $t_0$ and different local Alfv\'en velocities $v_{\rm A0}$. Thus, experimenting with all these different parameters demonstrated that the Moreton wave could not be reproduced when taking the kinematics of the radial outward movement of the CME as input for the model.

This finally pushed us to use synthetic kinematics in order to imitate other possible drivers for the signal. So far it was clearly derived from the model that the source expansion needs to be more impulsive (early shock formation).
For synthetic kinematics of stronger and shorter acceleration of the source surface expansion (3-D piston type) we found a good match between the model generated signal and the observed Moreton wave. The timing of the shock
formation is, when using these kinematical profiles, in good agreement with the appearance of the first Moreton wave front. Using an exponential attenuation factor (see Equ.~\ref{expon}) with short signal decay lengths the best match to the observed Moreton wave could be found. On average the Alfv\'en Mach number $M_{\rm A}$ from such synthetic kinematics are within the range of $M_{\rm A}\approx$1.5--3 which agrees with observed Alfv\'en Mach numbers for Moreton waves \citep[e.g.][]{narukage02,warmuth04a}. The initial source size and its expansion dimension that is necessary to mimic the observed Moreton wave can be interpreted as the laterally expanding CME flanks or the volume expansion of the flare.

\cite{pomoell08} concluded from a 2D magnetohydrodynamic simulations that for the driver of a Moreton wave a high acceleration during a short time interval is necessary. This was interpreted to require a strong lateral expansion, either lift-off of an over-pressured flux rope or thermal explosion-kind of energy release. Likewise, \cite{zic08} obtained from a 3D analytical model that a short acceleration phase up to high velocities ($\sim$1000~km~s$^{-1}$) within a low Alfv\'en velocity environment is necessary to create a shock that is capable of causing type II bursts in the dm/m wavelength range and H$\alpha$ Moreton waves.

Concluding, for the January 17, 2005 event under study it is unlikely that the bow shock of
the CME generated the observed Moreton wave. The CME is a too gradually accelerating
source in the lift-off phase and a too fast one in the later evolution phase to cause the observed Moreton wave kinematics. An impulsively accelerated expansion of a source surface acting as a temporary piston would be a more appropriate mechanism to generate the observed Moreton wave. Possible driving mechanisms would be the laterally expanding CME flanks or the impulsive volume expansion of the flare. The latter scenario would be in
accordance with the flare initiated ``blast wave'' scenario proposed from observational results for
the kinematics of Moreton waves \citep[see][]{warmuth01,warmuth04a,vrsnak02a} but in contrast to the
numerical model by \cite{chen02} who claimed that Moreton waves correspond to the piston-driven
shock over the CME. For the future it would be important to have more such complete data sets
including both, observations from the early CME evolution (upward moving front as well as expanding
flanks) and detailed observations of Moreton waves, in order to validate the presented results.

\acknowledgements M.T. is supported by the Austrian {\em Fonds zur F\"orderung der
wissen\-schaftlichen Forschung} (FWF Erwin-Schr\"odinger grant J2512-N02). T.{\v{Z}}. and B.V. acknowledge funding by the Croatian Ministry of Science, Education and Sports under the project 007-0000000-1362. A.V. gratefully acknowledges the Austrian {\em Fonds zur F\"orderung der wissen\-schaftlichen Forschung} (P20867-N16).

\bibliographystyle{apj}

\end{document}